\documentclass{article}
\usepackage{ijcai25}

\usepackage{amsmath,amsfonts,bm}

\def\eqref#1{equation~\ref{#1}}

\def\1{\bm{1}}

\def\rva{{\mathbf{a}}}

\def\rvh{{\mathbf{h}}}

\def\rvt{{\mathbf{t}}}

\def\rvx{{\mathbf{x}}}
\def\rvy{{\mathbf{y}}}
\def\rvz{{\mathbf{z}}}

\def\rmT{{\mathbf{T}}}

\DeclareMathAlphabet{\mathsfit}{\encodingdefault}{\sfdefault}{m}{sl}
\SetMathAlphabet{\mathsfit}{bold}{\encodingdefault}{\sfdefault}{bx}{n}

\usepackage{times}
\usepackage{soul}
\usepackage{url}
\usepackage[hidelinks]{hyperref}
\usepackage[utf8]{inputenc}
\usepackage[small]{caption}
\usepackage{graphicx}
\usepackage{amsmath}
\usepackage{amsthm}
\usepackage{booktabs}
\usepackage{algorithm}
\usepackage{algorithmic}
\usepackage[switch]{lineno}

\def\eg{\emph{e.g., }}
\def\ie{\emph{i.e., }}

\def\wrt{\emph{w.r.t. }}

\def\name{\textit{Inter3D}}
\newcommand{\bfsection}[1]{\vspace*{0.0mm}\noindent\textbf{#1.}}
\usepackage{xcolor}

\usepackage {amssymb}

\title{Inter3D: A Benchmark and Strong Baseline for Human-Interactive \\3D Object Reconstruction}

\author{
Gan Chen$^1$\and
Ying He$^{1,2}$\and
Mulin Yu$^3$\and
F. Richard Yu$^{1,2}$\and
Gang Xu$^2$\and
Fei Ma$^2$\and
Ming Li$^2$\And
Guang Zhou$^2$\\
\affiliations
$^1$Shenzhen University,
$^2$Guangdong Laboratory of Artificial Intelligence and Digital Economy (SZ),
$^2$Shanghai AI Laboratory
}

\begin{document}

\maketitle

\begin{abstract}
Recent advancements in implicit 3D reconstruction methods, \eg neural rendering fields and Gaussian splatting, have primarily focused on novel view synthesis of static or dynamic objects with \textit{continuous} motion states. However, these approaches struggle to efficiently model a human-interactive object with $n$ movable parts, requiring $2^n$ separate models to represent all \textit{discrete} states. To overcome this limitation, we propose \textit{\name{}}, a new benchmark and approach for \textit{novel state synthesis} of human-interactive objects. We introduce a self-collected dataset featuring commonly encountered interactive objects and a new evaluation pipeline, where only individual part states are observed during training, while part combination states remain unseen. We also propose a strong baseline approach that leverages Space Discrepancy Tensors to efficiently modelling all states of an object. To alleviate the impractical constraints on camera trajectories across training states, we propose a Mutual State Regularization mechanism to enhance the spatial density consistency of movable parts. In addition, we explore two occupancy grid sampling strategies to facilitate training efficiency. We conduct extensive experiments on the proposed benchmark, showcasing the challenges of the task and the superiority of our approach. The code and data are publicly available at \href{https://github.com/Inter3D-ui/Inter3D}{here}. The supplementary documents include more experimental results. 
\end{abstract}

\begin{figure}[!t]
  \centering
  \includegraphics[width=0.8\linewidth, keepaspectratio]{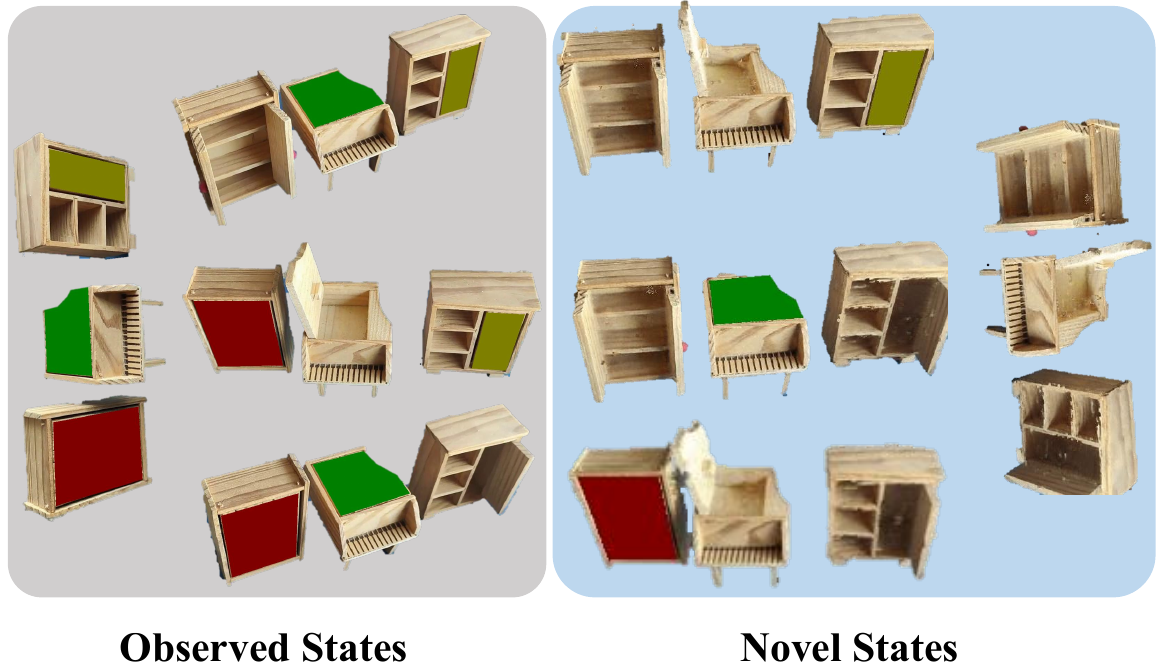}
  \vspace{-2mm}
   \caption{
    Illustration of reconstructing a human-interactive object, Furniture, in our \textit{\name{}}. The object consists of $n=3$ movable parts, highlighted in red, green, and olive-green. With $2^n=8$ \textit{discrete} states, the task requires significant computational and memory resources, making it impractical for existing methods to train separate models for each state. 
    Furthermore, ensuring consistency in external appearances and internal structures across states poses a significant challenge when states are trained independently.
    In contrast, our approach efficiently synthesizes \textit{novel combination states} by observing only the canonical and individual part states.
    }
  \label{fig:1}
  \vspace{-3mm}
\end{figure}

\section{Introduction}
3D object reconstruction has been a long-standing focus of research in computer vision and graphics~\cite{Berger:survey}. Recently, neural radiance fields (NeRFs)~\cite{nerf} pioneer volumetric rendering to enable high-fidelity modelling of 3D objects with fine-grained geometric structures, facilitating its widespread applications across various domains~\cite{mip360,kplane,dnerf,dreamfusion,genn2n}. 3D Gaussian splatting~\cite{3DGS} extends the principles of NeRFs by representing an object with a set of 3D Gaussian primitives and synthesizes novel views via rasterization, offering a more expressive and rendering-efficient 3D representations~\cite{dy3dgs}. 

Despite their advancements, existing methods predominantly focus on modeling static objects or dynamic ones with simple \textit{continuous} motions, where the states of objects exhibit no abrupt changes. Also, these methods primarily focus on synthesizing novel views of the observed states during training. However, many objects commonly encountered in daily life are interactive and have numerous \textit{discrete} states. A human-interactive object with \( n \) movable parts results in a total of \( 2^n \) states. Efficiently modeling such objects holds significant potential for applications in virtual reality~\cite{vr}, gaming~\cite{gaming}, and embedded AI~\cite{embodiedAI}.

As illustrated in Figure~\ref{fig:1}, the Furniture object consists of three movable parts, highlighted in red, green, and olive-green. It features one canonical state and three individual states where only one part is open, along with four combination states where two or more parts are open. Existing methods fail to efficiently model all possible states of the Furniture object. A straightforward solution would be to train a separate 3D model for each object state. However, this solution incurs prohibitive computational and memory overhead. Moreover, different states of the same human-interactive object inherently share consistent external appearances and internal structures, which this solution fails to capture.

To address these limitations, we propose \textit{\name{}}, a novel benchmark and method for human-interactive 3D object reconstruction. \textit{\name{}} introduces a self-collected dataset comprising commonly encountered yet challenging interactive objects in daily life, such as Car, Display Case, Drawer, and Furniture. It also introduces a new evaluation scenario, \ie \textit{novel state synthesis}, in which only the canonical and individual part states of an object are observed during training, while all combination states remain unseen and are reserved for evaluation, as illustrated in Figure~\ref{fig:1}. This setup is particularly challenging as no views of combination states are used during training, yet the synthesized novel states must remain consistent with the observed states.

Our method effectively tackles this challenging scenario by leveraging Space Discrepancy Tensors in combination with multi-resolution hash encoding (MHE)~\cite{ngp} to efficiently model all states of an interactive object. Specifically, MHE is initially employed to represent the canonical state of the object, where all movable parts remain untouched. The Space Discrepancy Tensors capture the inherent variations between each manipulated part state and the canonical state, providing a compact and expressive representation for the decomposition of individual movable parts. To mitigate rendering quality degradation caused by trajectory perturbations across training states, we introduce a Mutual State Regularization mechanism to ensure consistency of movable parts across states. Additionally, we explore two new occupancy grid sampling schemes to optimize training efficiency. After training on the canonical state and individual part states, our method can synthesize arbitrary combination states by selecting the maximum density difference.

Our key contributions in this work are summarized as follows:
\begin{itemize}
    \item 
    We introduce a novel benchmark for human-interactive 3D object reconstruction, supported by a self-collected dataset of commonly encountered interactive objects and a novel state synthesis testing scenario. This benchmark provides a foundation for advancing research in this field.
    
    \item 
    We propose a novel approach that integrates Space Discrepancy Tensors with multi-resolution hash encoding, enabling efficient modeling of all possible states of interactive objects. This approach provides a strong baseline for future studies.
    
    \item We develop a Mutual State Regularization mechanism to mitigate rendering artifacts caused by trajectory inconsistencies and design two occupancy grid sampling strategies to balance memory usage and training speed.
\end{itemize}

\begin{figure*}[!t]
 \centering
    \includegraphics[width=\linewidth]{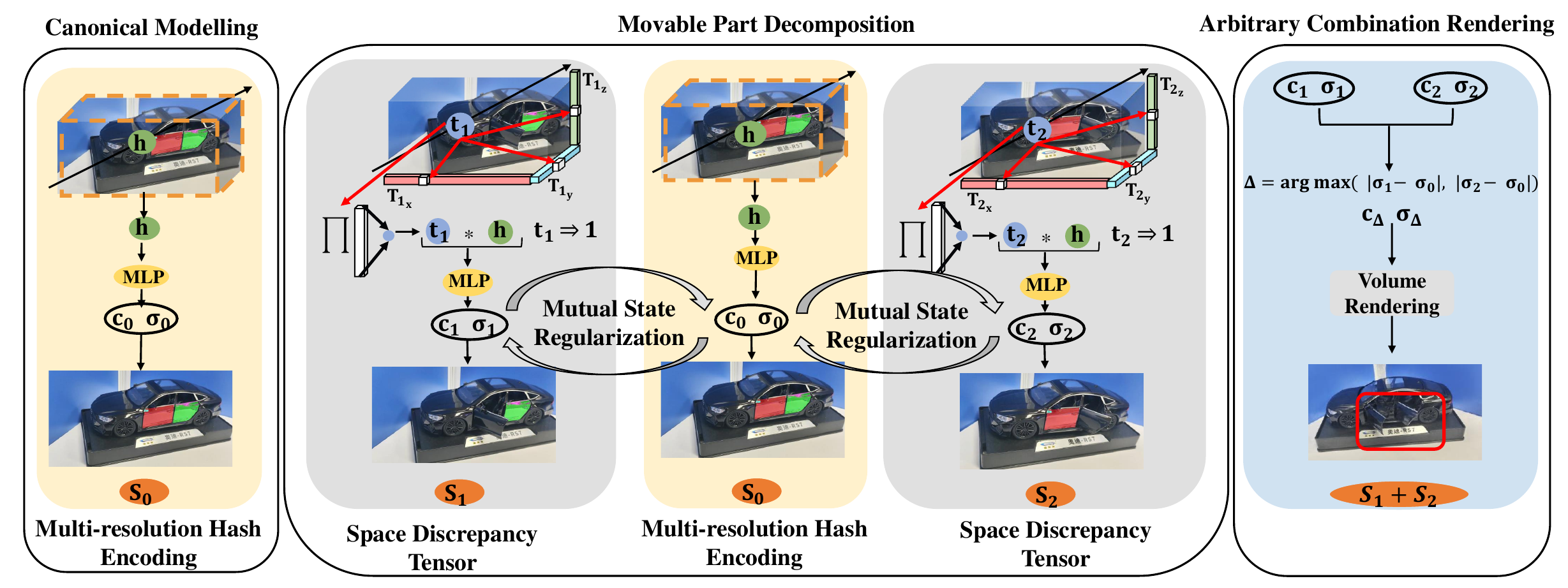}
     \vspace{-5mm}
    \caption{ 
    Overview of our approach, which comprises three key stages, \ie Canonical Modelling, Movable Part Decomposition, and Arbitrary Combination Synthesis. 
    The inactive movable parts in each state are covered by colored masks.
    In \textbf{Canonical Modelling}, the canonical state $S_0$ of the interactive object, with all $n$ movable parts closed, is reconstructed using InstantNGP. The spatial sample point features are retrieved via multi-resolution hash encoding, represented as $\rvh$, with attributes color $c_0$ and density $\sigma_0$ projected through a multi-layer perceptron (MLP). In \textbf{Movable Part Decomposition}, each movable part is manipulated sequentially, resulting in states where only one part is open, denoted as $S_i, i\in\{1,...,n\}$. For each $S_i$, the sample point features are represented as $\rvh * \rvt_i$, where $\rvt_i$ is derived from the proposed Space Discrepancy Tensors, encoding the differences between $S_i$ and $S_0$. To address camera trajectory perturbations across states, we introduce a Mutual State Regularization mechanism, which mitigates rendering artifacts caused by movable part  misalignment. In \textbf{Arbitrary Combination Synthesis}, to render the combined novel state $S_i+S_j$, the sample point features with the maximum density difference, selected from $\{|\sigma_i-\sigma_0|, |\sigma_j-\sigma_0|\}$, are used for volumetric rendering, which enables the efficient modelling of arbitrary combinations of movable parts. 
    }
  \label{fig:2}
  \vspace{-2mm}
\end{figure*}

\section{Related Works}
\label{sec:Related Work}
\bfsection{Static 3D Reconstruction}
Recent advances in implicit 3D reconstruction, particularly with neural radiance fields (NeRFs)~\cite{nerf}, have enabled high-quality novel view generation by predicting volume density and radiance from 3D coordinates and viewpoint directions. Variants like mip-NeRF~\cite{mip360} optimize performance in more complex scenarios. 3D Gaussian splatting~\cite{3DGS} uses a set of Gaussian primitives with learnable parameters, \ie positions, covariance matrices, opacity, and spherical harmonics coefficients, and differentiable tile rasterizer to improve synthesis quality and training speed. However, these methods primarily focus on modeling static objects and can only be applied to a single state of a human-interactive object. To model all possible states, a series of independent 3D models would be required, significantly increasing computational costs and failing to ensure consistency across states.


\bfsection{Dynamic 4D Reconstruction}
Dynamic 4D reconstruction methods, such as K-Plane~\cite{kplane}, D-NeRF~\cite{dnerf}, and 4D Gaussians~\cite{4dgaussian}, extend vanilla NeRFs or Gaussian splatting by incorporating motion factors to model geometric and texture changes of time-varying objects. 
These methods excel at modeling dynamic objects with continuous states and synthesizing novel views of the training states. However, our approach focuses on modeling human-interactive objects with numerous discrete states and emphasizes novel state synthesis, a task that is infeasible for these methods.


\begin{figure}[!t]
 \centering
  \includegraphics[width=\linewidth, keepaspectratio]{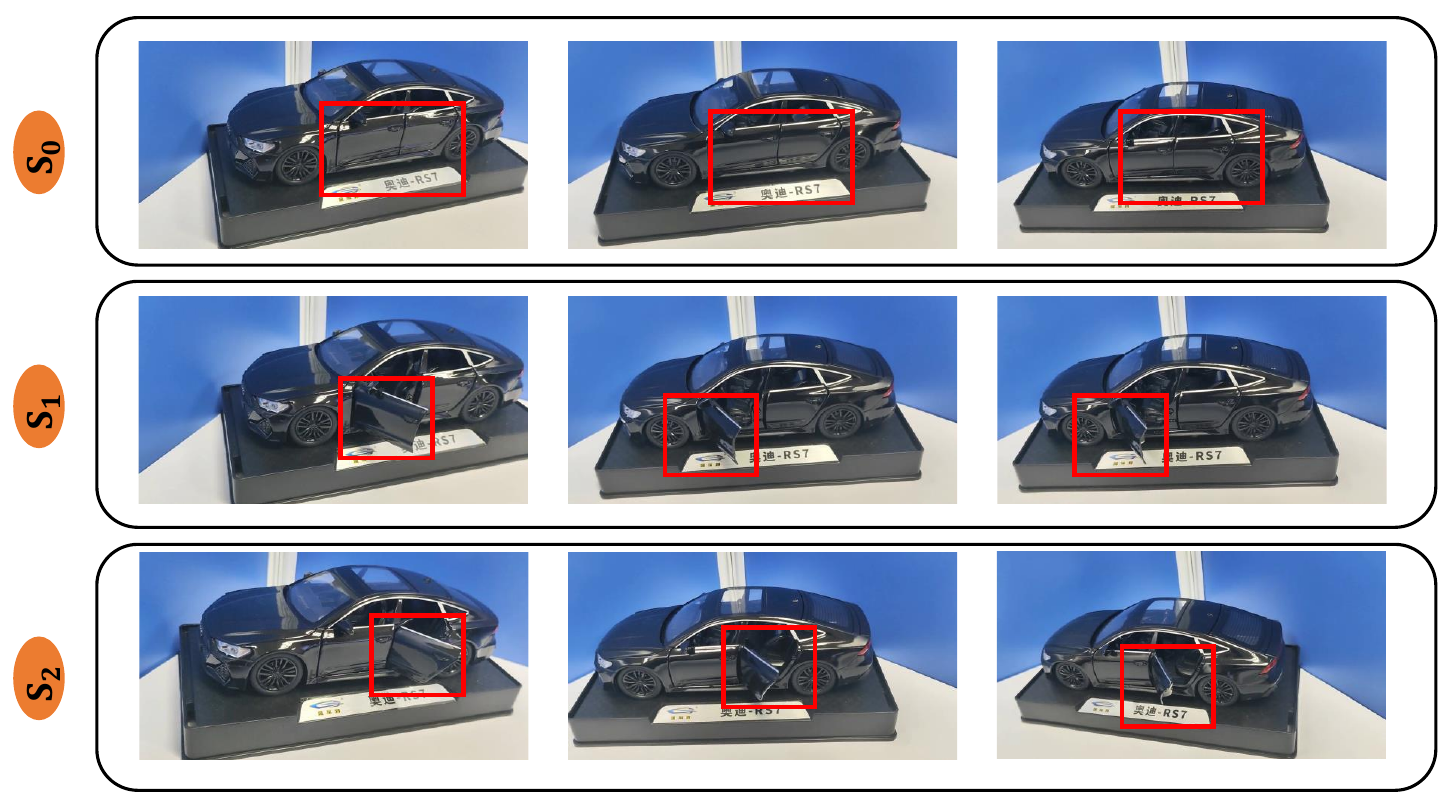}
   \vspace{-4mm}
    \caption{Data collection example of the object \textbf{Car} in our human-interactive benchmark \textit{\name{}}. A sequence of forward-facing images are captured for the canonical state $S_0$ with doors closed, the state $S_1$ with the front door open, and the state $S_2$ with the rear door open.}
  \label{fig:3}
  \vspace{-4mm}
\end{figure}

\section{Human-Interactive Benchmark}
\subsection{Data Collection}
To build the benchmark, we collect a diverse dataset comprising four objects with multiple movable parts commonly encountered in daily life: Car, Display Case, Drawer, and Furniture. These objects are arranged in increasing order of complexity. The \textbf{Car} represents smooth-surfaced objects with two movable parts, \ie front and rear doors, providing a straightforward yet realistic interaction scenario. The \textbf{Display Case}, with its two movable \textit{glass} doors and intricate internal contents, introduces additional complexity, especially for assessing the effectiveness of our approach under low-density conditions. The \textbf{Drawer}, featuring a typical three-drawer design commonly seen in daily environments, serves as a representative example of structured objects with discrete states. Finally, the \textbf{Furniture}, consisting of three types of cabinets, each with one movable part, features a challenging sandy background that introduces additional noise and interference, making it the most complex object in our dataset.

We capture object videos using handheld, forward-facing mobile shots and utilize SAM2 \cite{ravi2024sam2} to segment movable parts that remain in place. For each object, 60 images are extracted through uniform sampling at regular intervals to represent the canonical state and each individual state. As illustrated in Figure~\ref{fig:3}, we collect forward-facing data for a toy car with two movable parts, \ie its front and rear doors. Images are captured for the canonical state \(S_0\) and for individual part states \(S_1\) and \(S_2\), where only one door is opened at a time. Camera poses are extracted using COLMAP~\cite{colmap}, ensuring precise alignment for subsequent learning.

\subsection{Novel State Evaluation}
In traditional 3D or 4D reconstruction scenarios, objects are either static or exhibit \textit{continuous} states. Furthermore, models typically observe all possible states during training, with only novel views used for evaluation. In contrast, our benchmark introduces interactive objects with numerous \textit{discrete} states, where the evaluating states are novel and remain entirely unseen during training. 

For an object with \( n \) movable parts, there are \( 2^n \) possible states. Each part is manipulated individually, resulting in an individual state \( S_i, i \in \{1, \dots, n\} \), which is observed during training. When multiple parts are manipulated simultaneously, the resulting state becomes a \textit{spatial} combination of individual states. For instance, \( S_i + S_j, i, j \in \{1, \dots, n\} \), represents the simultaneous manipulation of the \( i \)th and \( j \)th parts. These combinatorial states are entirely unseen during training and reserved for evaluation rendering, significantly increasing the complexity of the task. This challenging setup requires models to achieve a higher level of generalization and adaptability, thereby pushing the boundaries of existing 3D reconstruction techniques.


\section{Human-Interactive 3D Reconstruction}
\textit{\name{}} establishes a new benchmark for synthesizing diverse states of human-interactive objects, offering a powerful baseline method. Unlike classical approaches, which require $2^n$ separate models to represent an object with $n$ movable parts, our method efficiently represents the object using a unified model. Figure~\ref{fig:2} illustrates the framework of our approach. In this section, we first describe the details of InstantNGP \cite{ngp} in Sec.~\ref{preliminary}. Then, we detail the initialization Canonical Modelling in Sec.~\ref{Canonical Modelling}, which defines the canonical state of a human-interactive object and provides a solid foundation for the subsequent steps. Sec.~\ref{Movable Part Decomposition} introduces the Space Discrepancy Tensors, which efficiently models discrepancies between individual part states and canonical state. Sec.~\ref{Arbitrary Combination Synthesis} presents the pipeline for representing novel combination states, and elaborates how the Mutual State Regularization alleviates the impractical constraints on camera trajectories across training states. Finally, Sec.~\ref{Training Efficiency} discusses two occupancy grid sampling strategies designed to enhance training efficiency.

\subsection{Preliminaries} 
\label{preliminary}
InstantNGP \cite{ngp} boosts up the rendering and training speed of NeRFs. It proposed a multi-resolution voxel grid, implicitly encoded using hash functions, enabling efficient optimization and rendering with compact models. A ray passing through three-dimensional space is defined as $\mathbf{r}(t) = \mathbf{o} + t\mathbf{d}$, where $\mathbf{o}$ is the ray's origin, $\mathbf{d}$ is its direction, and $t$ represents the distance along the ray. InstantNGP employs a density occupancy grid with step-based indexing to acquire sampling points. Then, based on the 3D coordinates of a sampling point, feature vectors of multiple surrounding voxels are retrieved from the hash table. These voxel feature vectors are then linearly interpolated to obtain the feature vector $\rvh$ for the sampling point. An MLP is used to compute the color $c$ and density $\sigma$ of the sampling points. The pixel color $C$ is determined by the following volume rendering equation applied to the $N$ sampling points along the ray:

{\small
 \vspace{-2mm}
\begin{align}
  C(\mathbf{r}) & = \int^{t_f}_{t_n} T(t)\sigma(\mathbf{r}(t))\mathbf{c}(\mathbf{r}(t),\mathbf{d})\ dt \ {\rm with},   \nonumber \\
  T(t) & =exp(-\int^{t}_{t_n}\sigma(\mathbf{r}(t^\prime))\ dt^\prime).
  \label{eq:1}
\end{align}
}
$\mathcal{L}_{MSE}$ is employed to minimize the discrepancy between the predicted pixel color and the ground truth $C^{gt}$:

{\small
\begin{align}
  \mathcal{L}_{MSE} =  \mathbb{E} \begin{bmatrix}\   || C - C^{gt} ||^2_2 \  \end{bmatrix}.
  \label{eq:2}
\end{align}
}

\subsection{Canonical Modelling}
\label{Canonical Modelling}
Canonical Modelling reconstructs the canonical state of the object, providing a solid foundation for the subsequent movable part decomposition, where a part of the object is manipulated into an individual state. As shown in Figure~\ref{fig:2}, we leverage InstantNGP to model the canonical state $S_0$.

To enhance the compactness of spatial points and reduce the presence of fuzzy clouds in space, we employ the distortion loss function from MipNeRF360~\cite{mip360}:

{\small
\begin{align}
    \mathcal{L}_\mathrm{dist}(s,w) 
    & = \sum_{i,j} w_i w_j 
    \Bigg\lvert
    \frac{s_i+s_{i+1}}{2} -\frac{s_j+s_{j+1}}{2} 
    \Bigg\rvert
    + \nonumber \\
    & 
    \ \frac{1}{3}\sum_i {w_i}^2\delta_i,
    \label{eq:3}
\end{align}
}

\noindent
where $w_i = (1 - e^{-\sigma_i\delta_i}) T_i$, and $s$ represents the distance from the sample point to the ray origin. Additionally, we incorporate a loss term aimed at minimizing the entropy of the cumulative weights, which helps to concentrate the distribution of spatial sample points:
{\small
\begin{align}
    \mathcal{L}_\mathrm{opacity} & = \mathbb{E} 
    \begin{bmatrix}
        - O \log{O}
    \end{bmatrix},
    \label{eq:4}
\end{align}
}

\noindent
where $O$ is calculated as the sum of the weights $w_i$ of $N$ sample points along the ray, given by:$\begin{smallmatrix} O &= \sum\limits_{i}^{N} w_i \end{smallmatrix}$.

\subsection{Movable Part Decomposition}
\label{Movable Part Decomposition}
In this section, we aim to identify the movable components from the interactive object, and learn their individual states. As illustrated in Figure~\ref{fig:2}, we train the model across all individual states, starting with the initialized canonical state $S_0$. Instead of explicitly identifying the moving parts, we propose Space Discrepancy Tensors to implicitly encode the individual states, enhancing the scalability and flexibility of our method.

Specifically, the Space Discrepancy Tensors are three spatially correlated feature tensors in the Cartesian coordinate system, denoted as $\rmT_{i_\rvx}$, $\rmT_{i_\rvy}$, and $\rmT_{i_\rvz}$, each with a shape of 512$\times$32, where 512 represents the spatial resolution and 32 denotes the feature dimension. These tensors are leveraged to capture the discrepancies introduced by the individual state $S_i$. We normalize the 3D coordinates of a random sampling point $p$ and project them onto these three feature tensors. Using nearest-neighbor sampling, we extract the spatially relevant features along each axis, and then compute the final feature vector $\rvt_i$ of $p$ by element-wise multiplication:

{\small
\begin{align}
    \rvt_i = \prod_{\rva\in\{\rvx,\rvy,\rvz\}} \psi(\rmT_{i_\rva},\pi_\rva(p)),
    \label{eq:5}
\end{align}
}

\noindent
where $\pi_\rva(p)$ represents projecting $p$ onto the axis $\rva$, and $\psi(\rmT_{i_\rva},\cdot)$ denotes the nearest-neighbor sampling of a point feature from the tensor $\rmT_{i_\rva}$.
To ensure the non-interactive part consistency between the canonical state and individual states, similar to K-Planes~\cite{kplanes}, we employ $\mathcal{L}_{consis}$ to constrain $\rmT_i$ towards 1:

{\small
\begin{align}
    \mathcal{L}_{consis}(\rmT_i) = \sum_{\rva\in \{\rvx,\rvy,\rvz\}} ||1-\rmT_{i_\rva} ||_1.
    \label{eq:6}
\end{align}
}We also employ the Laplacian (second-order derivative) filter $\mathcal{L}_{smooth}$ in K-Planes~\cite{kplanes} to encourage the smooth variations in the state $S_i$.



\subsection{Arbitrary Combination Synthesis}
\label{Arbitrary Combination Synthesis}
\begin{figure}[tb]
 \centering
  \includegraphics[width=\linewidth, keepaspectratio]{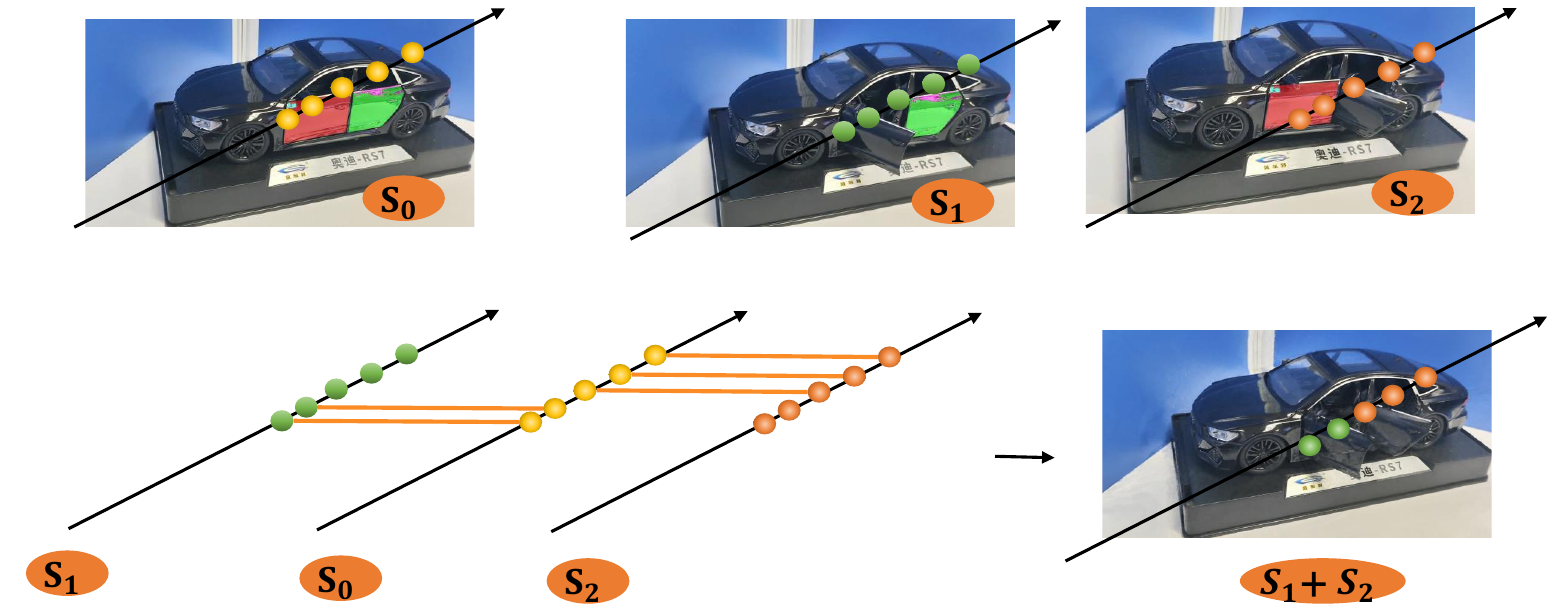}
    \caption{Illustration of synthesizing the novel combination state on the \textbf{Car} object. Given the features of the same sample point across the training states, its optimal feature on the novel state is determined by comparing the individual state $S_i$ with the canonical state $S_0$. The inactive movable parts in each state are covered by colored masks.}
  \label{fig:4}
  \vspace{-4mm}
\end{figure}

We propose a simple yet effective strategy of synthesizing a novel combination state on top of the training states. For example, the state $S_1+S_2$ in Figure~\ref{fig:4} can be generated by spatially merging $S_1$ and $S_2$. Specifically, our goal is to combine the varying components of $S_1$ and $S_2$ relative to $S_0$, along with the non-interactive parts of $S_0$, to form a new state $S_1 + S_2$. The sample point density $\sigma_i$ associated with the individual state $S_i$ are employed to capture the differences \wrt the canonical state $S_0$ and the state with the maximum differences dominates the volume rendering:
{\small
\begin{align}
    \Delta = \arg \max_i\ |\sigma_i-\sigma_0| \ \ \ i \in \{1,\dots,n\} .
    \label{eq:7}
\end{align}
}The corresponding density and color $c_\Delta$ and $\sigma_\Delta$ are then queried to render the new state $S_1 + S_2$ via Equation~\ref{eq:1}. It is straightforward to adapt this synthesis strategy to an interactive object with more movable parts.

\bfsection{Mutual State Regularization}
In practice, it is infeasible to maintain absolutely identical camera trajectories across different training states, which degrades the synthesis quality of movable parts. To address this issue, we propose the Mutual State Regularization (MSR) to enhance the spatial consistency of movable parts between training states.

\begin{figure}[htb]
 \centering
  \includegraphics[width=\linewidth, keepaspectratio]{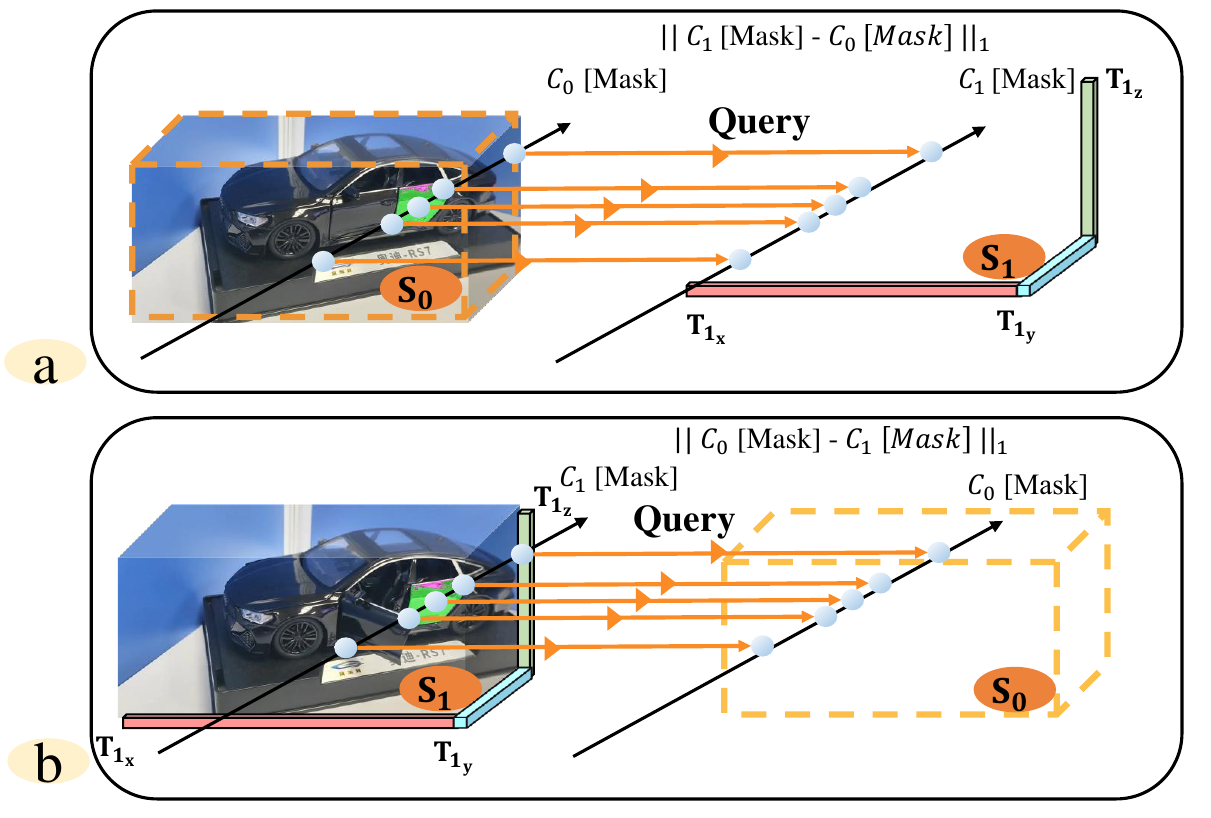}
    \caption{Illustration of our Mutual State Regularization mechanism on the \textbf{Car} object. MSR regularizes the sampling point features of the inactive movable part in the individual state $S_i$ with the canonical state $S_0$  through pixel alignment, and vice versa. $\mathcal{L}_1$ loss is applied to ensure consistency.}
  \label{fig:5}
  \vspace{-4mm}
\end{figure}

As shown in Figure~\ref{fig:5}.a, we match the inactive movable part, \ie the back door, of $S_1$ to that of $S_0$. Specifically, we randomly sample a set of points along a ray passing through the back door of $S_0$ (segmented by SAM2~\cite{sam2}) and obtain the pixel value $C_0[\text{Mask}]$ via integration. We then project the points onto the $S_1$ space and query the pixel value $C_1[\text{Mask}]$. The $\mathcal{L}_{MSR}$ loss is employed to align them, ensuring the back door consistency: 

{\small
\vspace{-10pt}
\begin{align}
  \mathcal{L}_{MSR} =  || C_1[Mask] - C_0[(Mask)] ||_1.
  \label{eq:8}
\end{align}
}Reversely, in Figure~\ref{fig:5}.b, we align the pixel values of the back door of $S_0$ with those of $S_1$. The alignment process can be easily generalized to other individual state $S_i$.

\subsection{Training Efficiency}
\label{Training Efficiency}
As aforementioned, in Canonical Modelling, we employ InstantNGP to reconstruct the canonical state of an interactive object. In the Movable Part Decomposition stage, we model various individual states, each sharing non-interactive parts but featuring different movable parts. To achieve this, we propose two occupancy grid sampling strategies that balance training memory and speed.


\bfsection{Shared Occupancy Grid}
We reuse the occupancy grid from Canonical Modelling by only increasing its learned density values, which represent the manipulated movable parts in each individual state. This strategy saves the memory usage, but experimentally increases the number of sampling points in each ray, which slows down the training process.

\bfsection{Independent Occupancy Grid}
In this strategy, we set independent occupancy grids for individual states, initialized with the canonical state, and update them separately. While this approach significantly increases memory usage, it greatly enhances training speed by reducing the number of sampling points.

With these strategies, we can choose the better option depending on computational resources and preferences.

\subsection{Overall Objectives}
\label{Overall Objectives}
The first two stages of our approach involves training. The learning objective in Canonical Modelling $\mathcal{L}_{CM}$ is as follows:

{\small
\vspace{-10pt}
\begin{align}
    \mathcal{L}_{CM} = \lambda_1\mathcal{L}_\mathrm{dist} + \lambda_2\mathcal{L}_\mathrm{opacity} + \mathcal{L}_{MSE},
    \label{eq:9}
\end{align}
}

\noindent
where $\lambda_1$ and $\lambda_2$ are the weighting coefficients, both set to 1e-3 in our experiments.
In Movable Part Decomposition, an individual state $S_i$ is randomly selected for training. We employ both $\mathcal{L}_{CM}$ and the learning objective $\mathcal{L}_{MPD}$ to optimize the model:

{\small
\vspace{-10pt}
\begin{align}
    \mathcal{L}_{MPD} = \lambda_3\mathcal{L}_{consis} + \lambda_4\mathcal{L}_{smooth} + \lambda_5\mathcal{L}_{MSR},
    \label{eq:10}
\end{align}
}

\noindent
where $\lambda_3$, $\lambda_4$ and $\lambda_5$ are set as 1e-4, 1e-3, and 1e-2, respectively.

\section{EXPERIMENTS}
\subsection{Implementation Details}
We employ the Adam optimizer~\cite{adam} with hyperparameters $lr=$ 1e-2, $\beta_1 = 0.9$, $\beta_2 = 0.99$, and $\epsilon=$1e-15 to train our model. The image resolution in our dataset is 1280$\times$680. Each epoch consists of 1,000 iterations and 8,192 pixels are randomly selected for each iteration. The Canonical Modelling stage requires two epochs to reconstruct the canonical shape of an interactive object. During the Movable Part Decomposition stage, a random individual state is selected for each iteration to learn the corresponding movable part. This phase takes 13 epochs. All experiments are conducted on a single NVIDIA GeForce RTX 3090 GPU.


\subsection{Case Study}
\begin{figure}[htb]
 \centering
  \includegraphics[width=\linewidth, keepaspectratio]{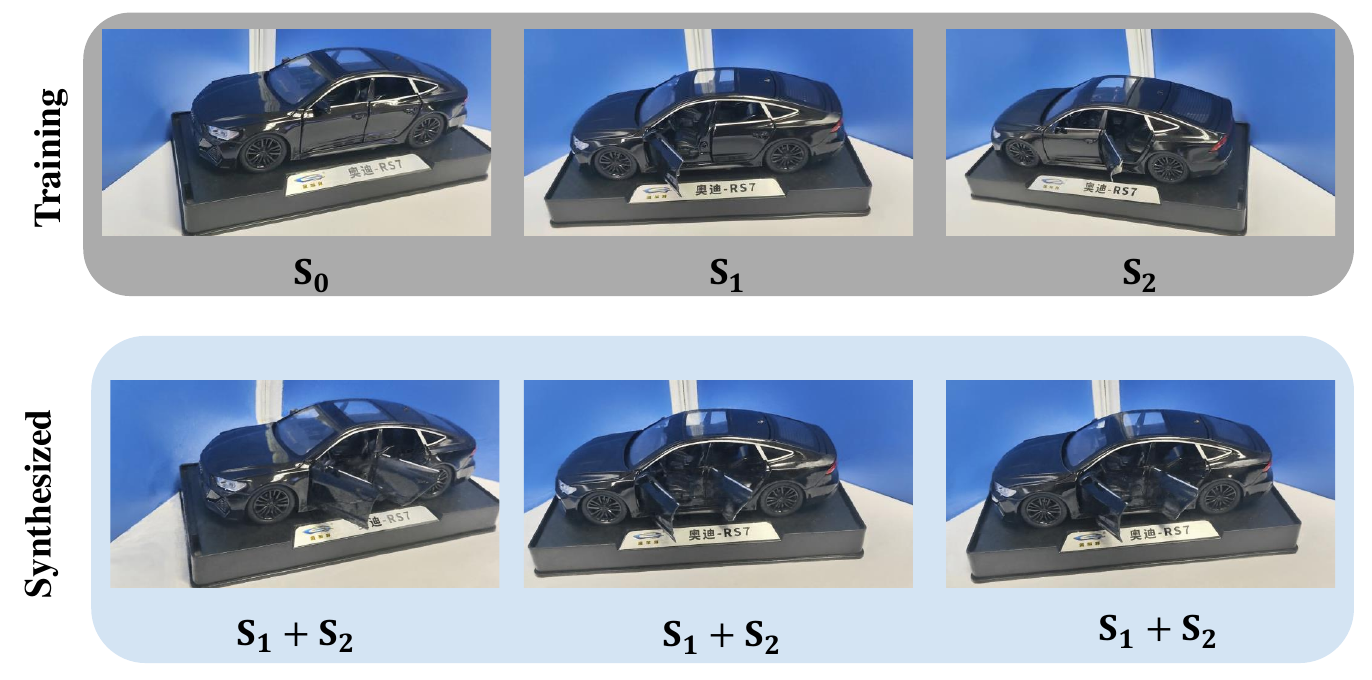}
    \caption{Training and synthesized states for the Car object in \textit{\name{}}. The canonical state and the individual states of two movable parts are observed during training. Our method successfully synthesizes various views of the novel combination state.}
  \label{fig:6}
  \vspace{-4mm}
\end{figure}

\bfsection{Car} As shown in Figure~\ref{fig:6}, the training states of the Car object consist of the canonical state $S_0$, the front door open state $S_1$, and the rear door open state $S_2$. The combination state $S_1 + S_2$, where both doors are open simultaneously, is entirely synthesized. The view visualizations showcase our efficacy.

\begin{figure}[htb]
 \centering
  \includegraphics[width=\linewidth, keepaspectratio]{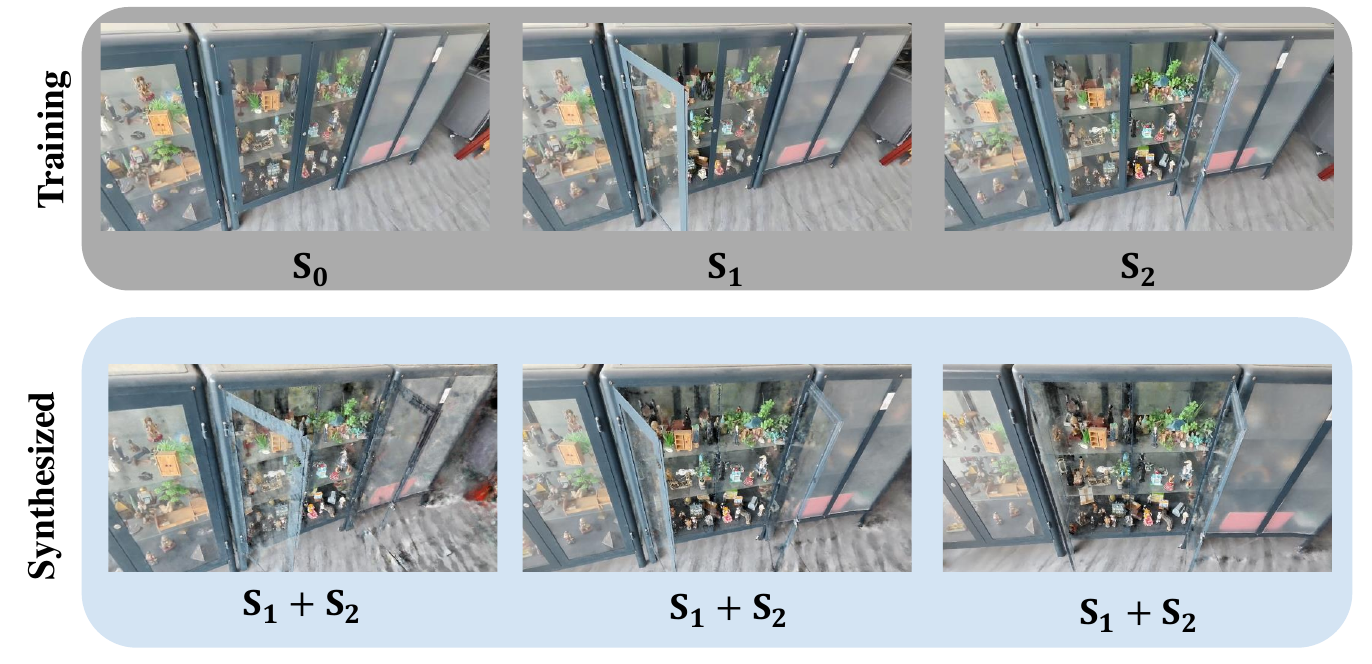}
    \caption{Training and synthesized states for the Display Case in \textit{\name{}}. Clearly, our method effectively handle the transparent object containing intricate gadgets.}
  \label{fig:9}
  \vspace{-4mm}
\end{figure}

\bfsection{Display Case}
Reconstructing a transparent object is inherently challenging, even without interactive components. The Display Case, illustrated in Figure~\ref{fig:9}, features two glass doors as movable components and contains numerous small gadgets, further complicating the scenario. Synthesized views from various poses demonstrate the effectiveness of our method in handling this complex setting.

\begin{figure}[htb]
 \centering
  \includegraphics[width=\linewidth, keepaspectratio]{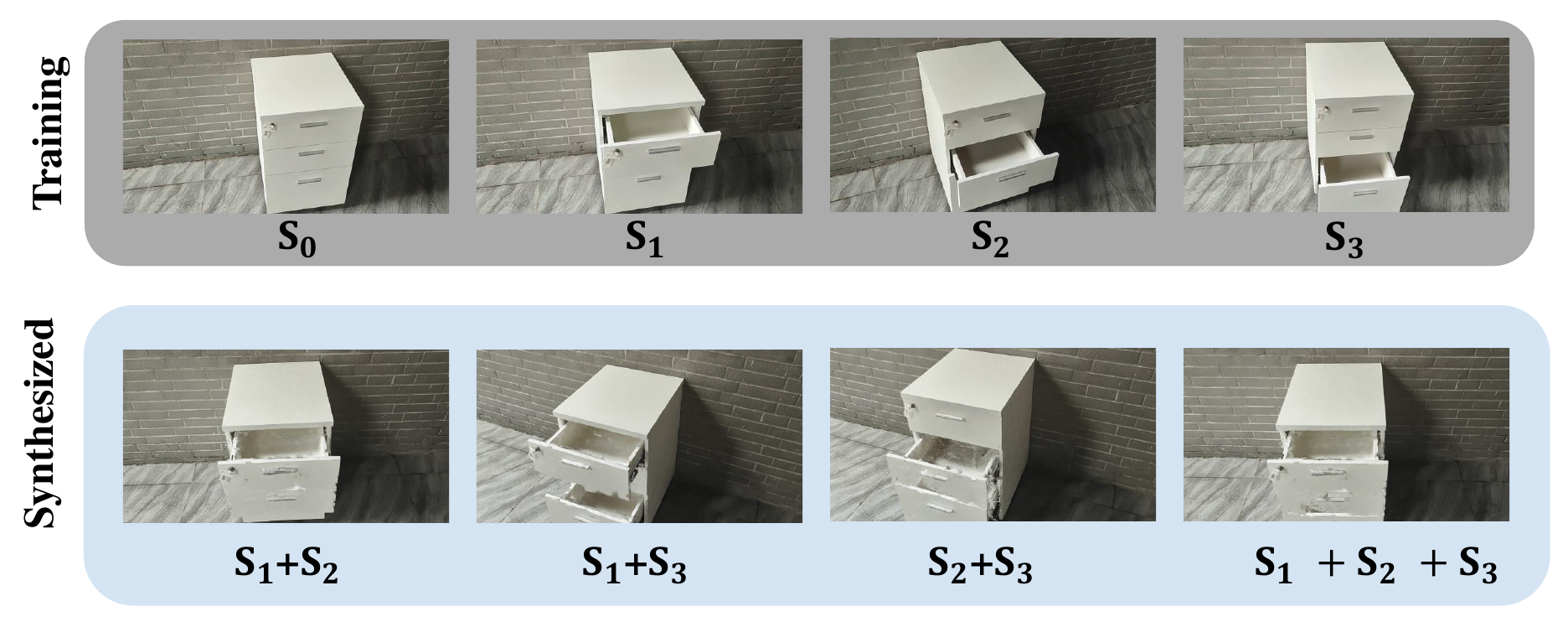}
    \caption{Training and synthesized states for the Drawer in \textit{\name{}}. Clearly, our method effectively model the complex object with more stacked movable parts.}
  \label{fig:8}
  \vspace{-4mm}
\end{figure}

\bfsection{Drawer}
The Drawer object, depicted in Figure~\ref{fig:8}, consists of three vertically arranged drawers, resulting in four novel combination states. The synthesized views across different states demonstrate consistent internal structures and external appearances.

\begin{figure}[htb]
 \centering
  \includegraphics[width=\linewidth, keepaspectratio]{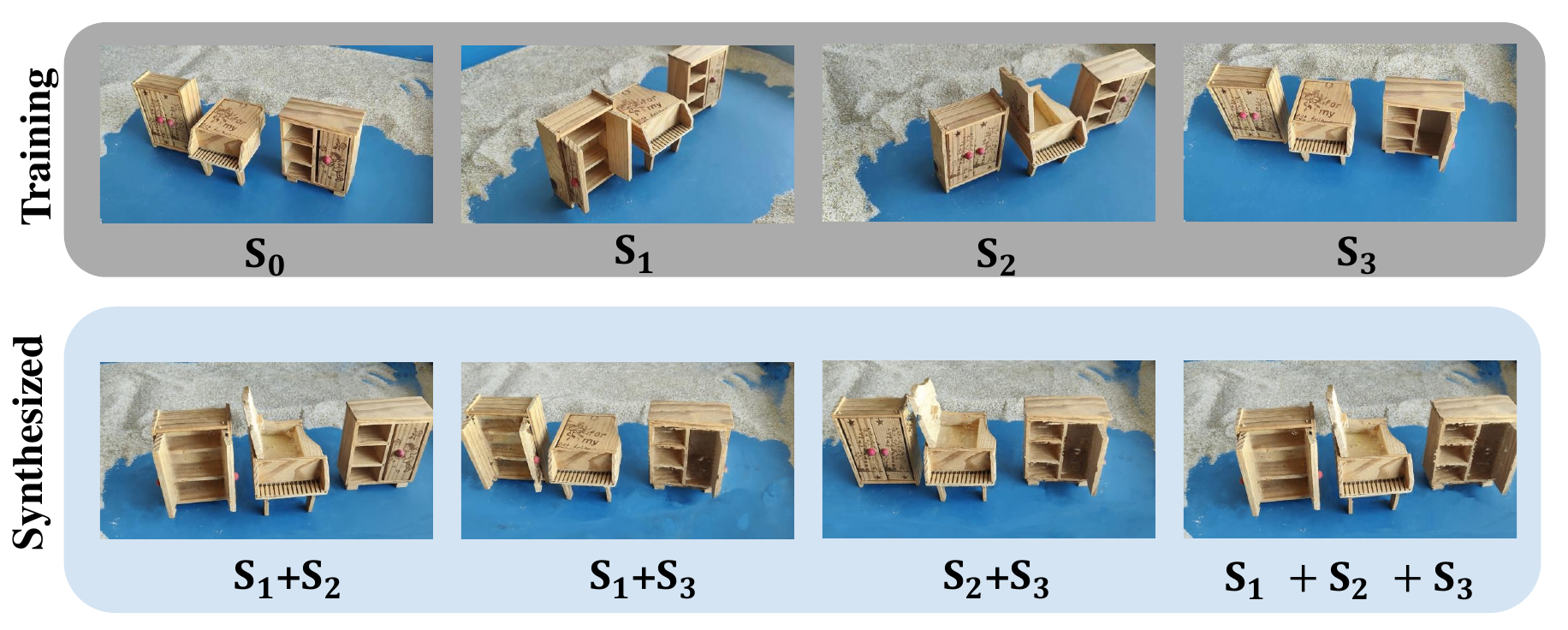}
    \caption{Training and synthesized states for the Furniture in \textit{\name{}}. The results demonstrate that our method can effectively model a highly challenging scene featuring three types of cabinets and a cluttered background.}
  \label{fig:7}
  \vspace{-2mm}
\end{figure}

\begin{figure}[htb]
 \centering
 \vspace{-2mm}
  \includegraphics[width=\linewidth, keepaspectratio]{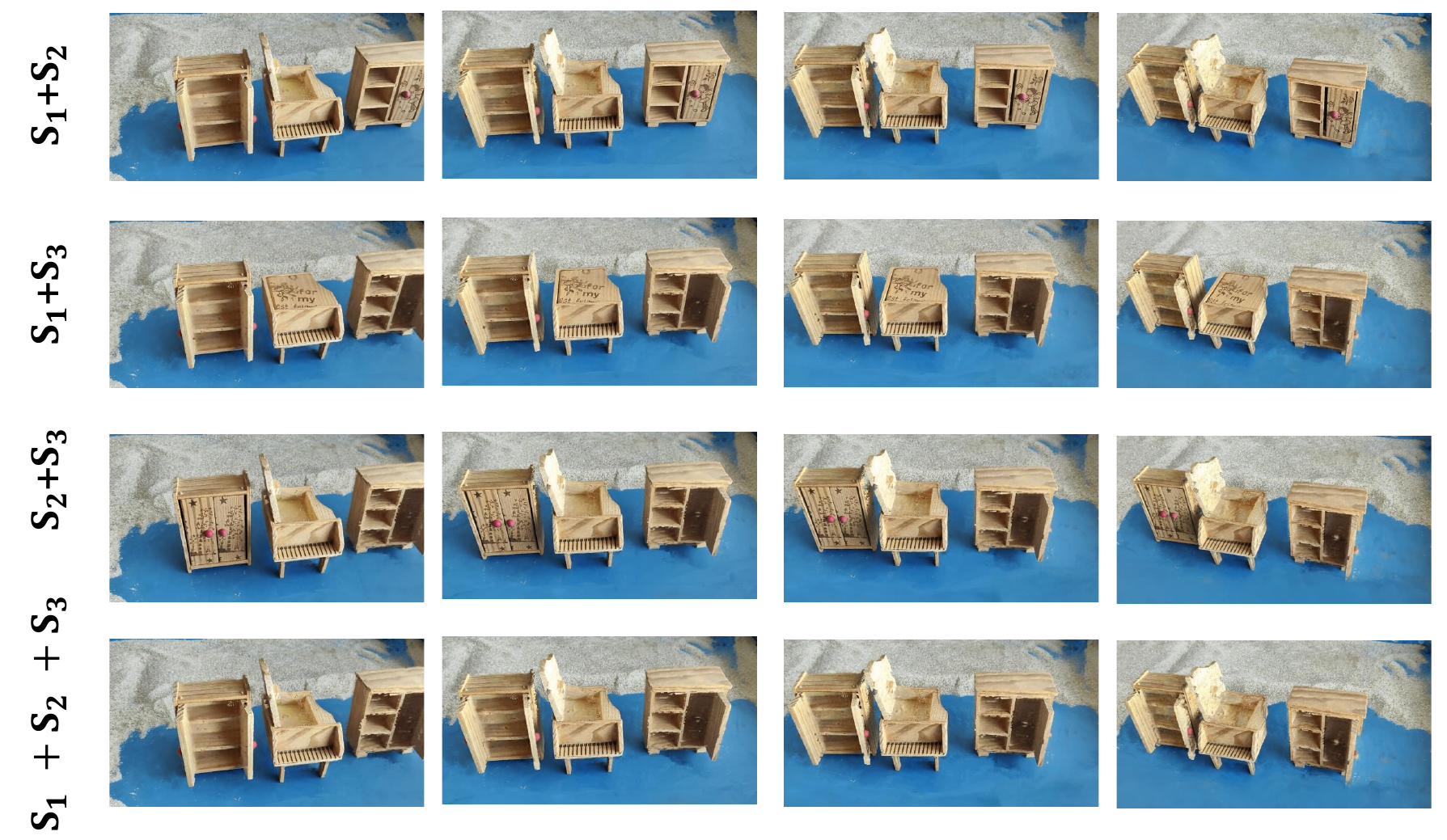}
    \caption{
    Synthesized combination states of the Furniture object in \textit{\name{}} under varying viewpoints. The visualizations demonstrate that our method produces realistic and geometrically consistent novel states across different views.
    }
  \label{fig:13}
\end{figure}

\bfsection{Furniture} Figure~\ref{fig:7} illustrates the Furniture scene, which includes three types of cabinets. Each cabinet features a movable part located on the front or the top and is surrounded by a pile of sand. The synthesized combination states demonstrate realistic cabinet appearances and consistent background rendering. Additionally, Figure~\ref{fig:13} presents rendered images from varying camera poses of multiple synthesized combination states, demonstrating that our method effectively generates geometrically consistent arbitrary novel states.

\begin{table*}[htb]
\center
\resizebox{0.8\linewidth}{!}{
\begin{tabular}{cccccccccc}
\toprule
\multicolumn{2}{c}{Scene}               & \multicolumn{2}{c}{Car}               & \multicolumn{2}{c}{Displayer Case}    & \multicolumn{2}{c}{Drawer}            & \multicolumn{2}{c}{Furniture}         \\
\multicolumn{1}{c|}{Strategies} & Metric & 
MA/MB & TT/s & 
MA/MB & TT/s & 
MA/MB & TT/s & 
MA/MB & TT/s  \\ \midrule
\multicolumn{2}{c}{Independent}         
& 318.28              & \textbf{969.44}          
& 318.28              & \textbf{1\,032.37}         
& 335.03              & \textbf{939.98}          
& 335.03              & \textbf{1\,120.48}         \\
\multicolumn{2}{c}{Share}               
& \textbf{284.78}              & 1\,264.99        
& \textbf{284.78}             & 1\,717.74         
& \textbf{285.53}              & 1\,386.01         
& \textbf{285.53}              & 1\,627.35         \\ \bottomrule
\end{tabular}}
\caption{Ablation study on the proposed occupancy grid sampling strategies. Across all scenes, the Independent strategy significantly reduces training time, while the Shared strategy requires considerably less memory. The best results are highlighted in bold.
}
\label{tab:2}
\vspace{-3mm}
\end{table*}

\subsection{Comparison with State-of-the-Arts}
\begin{figure}[htb]
 \centering
  \includegraphics[width=\linewidth, keepaspectratio]{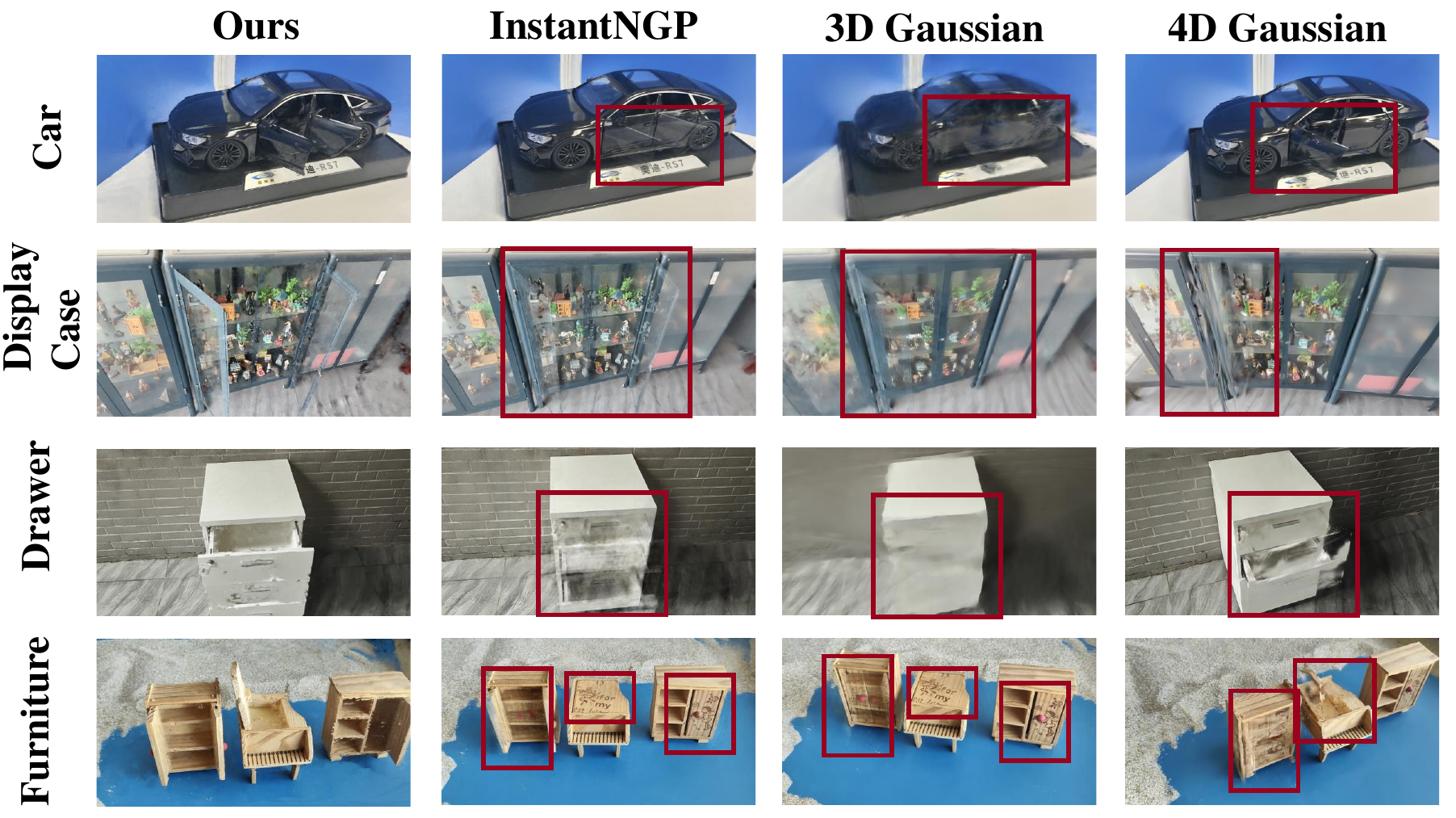}
    \caption{Comparison between our method and typical 3D and dynamic 4D object reconstruction methods. The highlighted regions demonstrate that existing methods fail to generalize to novel combination states, whereas our method successfully generates high-fidelity images of unseen states.
    }
  \label{fig:10}
  \vspace{-1mm}
\end{figure}

As the first work focusing on the novel state synthesis of a human-interactive object, we compare our method with three typical counterparts for 3D or 4D reconstruction, \ie InstantNGP~\cite{ngp}, 3D Gaussian~\cite{3DGS}, and 4D Gaussian~\cite{4dgaussian}, as shown in Figure~\ref{fig:10}. All methods are trained on the same states as ours, \ie the canonical state and individual states. For our method, we present the synthesized combination state where all movable parts are manipulated. For existing methods, we select the closest view to the combination state for comparison, as they lack the inherent designs of synthesizing novel states.

The highlighted visualizations in Figure~\ref{fig:10} reveal that existing methods fail to generate unseen states during training. For instance, in the Car object, InstantNGP simply overlays the movable parts, \ie the front and rear doors, while 3D Gaussian blends the doors together, failing to “open the doors” and resulting in visible quality degradation and prominent artifacts. Similarly, 4D Gaussian associates each individual state with time but still suffers from state overlap. In contrast, our method successfully renders the combination state where both car doors are open, and produce much higher-quality images.
%
%
%
In the Drawer object, InstantNGP reconstructs the background wall but fails to model the open drawers. 3D Gaussian is unable to reconstruct the scene entirely, missing both the environment and the open drawers. 4D Gaussian, on the other hand, produces significant ghosting artifacts in the drawers.
In comparison, our method successfully synthesizes the novel state with all drawers open. 
%
%
%
Furthermore, with the effective Movable Part Decomposition and Mutual State Regularization, our method achieves superior rendering quality on novel combination states.

\subsection{Ablation Study}
\begin{figure}[!t]
 \centering
  \includegraphics[width=\linewidth, keepaspectratio]{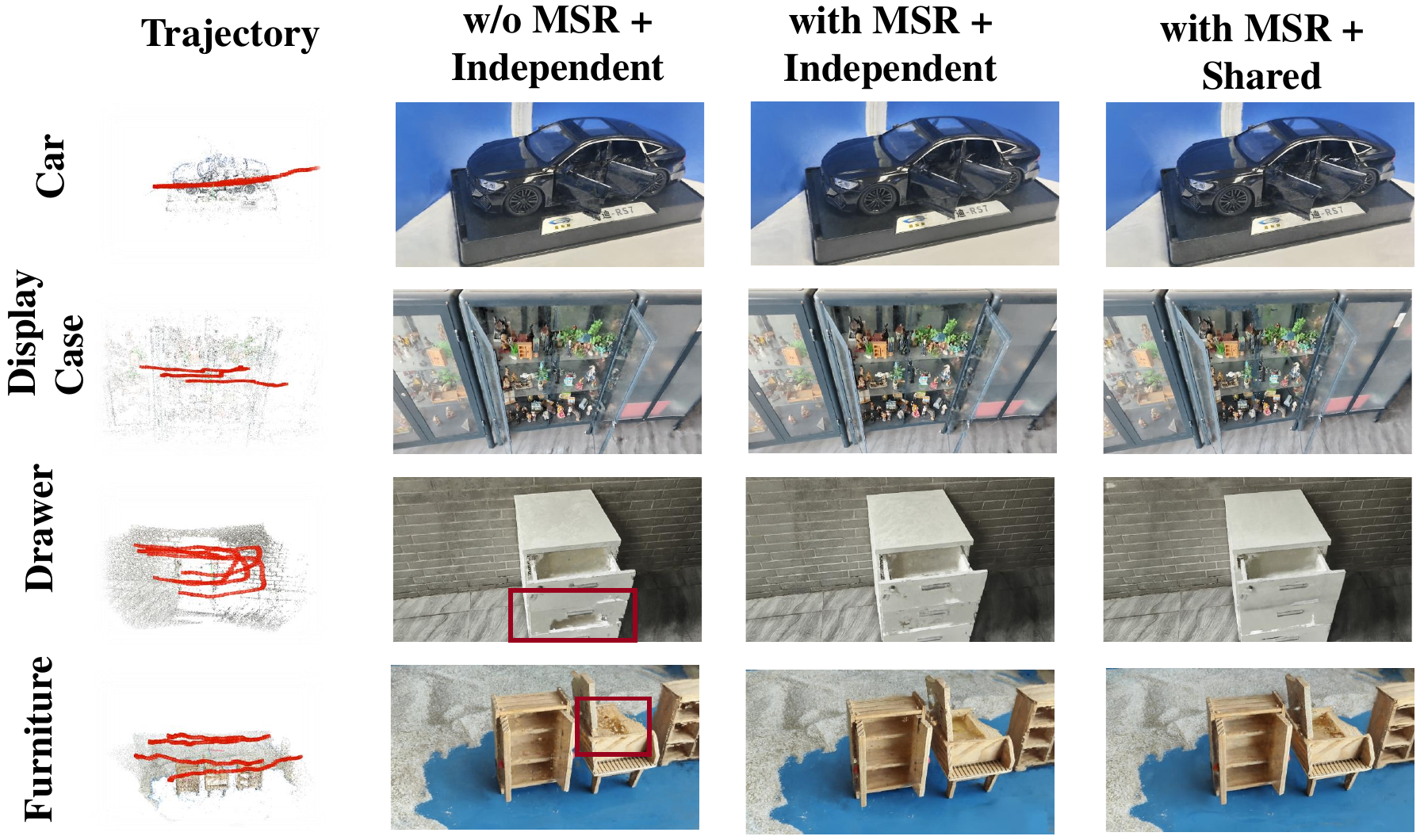}
    \caption{
    Ablation study on the proposed Mutual State Regularization and occupancy grid sampling strategies. The first column illustrates the camera trajectory perturbations across training states, which significantly degrade modeling efficacy, particularly for interactive objects with more movable parts. In contrast, the MSR mechanism effectively mitigates these negative effects. Additionally, the last two columns demonstrate that both the Shared and Independent Occupancy Grid sampling strategies perform very well.
    }
     
  \label{fig:12}
  \vspace{-1mm}
\end{figure}

As shown in Figure~\ref{fig:12}, we conduct ablation experiments on the proposed Mutual State Regularization (MSR) and occupancy grid sampling strategies.

\bfsection{Effect of Mutual State Regularization} 
The first column shows the camera trajectories used to capture training state images. The next two columns compare the synthesized novel states under the Independent Occupancy Grid. Evidently, trajectory inconsistencies lead to severe modeling failures, especially for objects with multiple movable parts, \ie Furniture and Drawer. Without MSR, noticeable artifacts appear in the highlighted regions of the Furniture, while missing regions are observed in the Drawer object.

\bfsection{Effect of Occupancy Grid Sampling Strategies}
The last two columns of Figure~\ref{fig:12} present the ablation results for the Independent and Shared Occupancy Grid sampling strategies. No significant differences in rendering quality are observed, demonstrating that both strategies are well-suited for our task. A detailed comparison of these strategies in terms of Memory Allocation (MA) and Training Time (TT) is provided in Table~\ref{tab:2}. Across all objects, the Independent strategy significantly reduces training time but requires more memory, while the Shared strategy exhibits the opposite behavior.

\section{Conclusion}
In this work, we present \textit{\name{}}, a comprehensive benchmark and a strong baseline method for \textit{novel state synthesis} of human-interactive objects with multiple movable parts. To facilitate this research, we construct a diverse dataset comprising commonly encountered but challenging interactive objects from everyday life. On this dataset, we introduce a novel and practical evaluation scenario, \ie novel state synthesis, where only canonical and individual part states are observed during training, while combination states are withheld for evaluation. To efficiently model the full range of states of an interactive object, we leverage Space Discrepancy Tensors with multi-resolution hash encoding, enabling precise learning of individual states. To ensure movable part consistency across training states, even under trajectory perturbations, we propose a Mutual State Regularization mechanism. Furthermore, we explore two occupancy grid sampling schemes to balance memory efficiency and training speed. Extensive experiments demonstrate that our approach achieves high-fidelity novel state synthesis, significantly outperforming existing methods. These results establish a strong foundation for advancing research in interactive 3D object reconstruction and open new avenues for exploring 3D object synthesis.

\newpage
\bibliographystyle{named}
\bibliography{ijcai25}

\begin{thebibliography}{}

\bibitem[\protect\citeauthoryear{Barron \bgroup \em et al.\egroup }{2022}]{mip360}
Jonathan~T Barron, Ben Mildenhall, Dor Verbin, Pratul~P Srinivasan, and Peter Hedman.
\newblock Mip-nerf 360: Unbounded anti-aliased neural radiance fields.
\newblock In {\em Proceedings of the IEEE/CVF conference on computer vision and pattern recognition}, pages 5470--5479, 2022.

\bibitem[\protect\citeauthoryear{Berger \bgroup \em et al.\egroup }{2017}]{Berger:survey}
Matthew Berger, Andrea Tagliasacchi, Lee~M Seversky, Pierre Alliez, Gael Guennebaud, Joshua~A Levine, Andrei Sharf, and Claudio~T Silva.
\newblock A survey of surface reconstruction from point clouds.
\newblock In {\em Computer graphics forum}, volume~36, pages 301--329. Wiley Online Library, 2017.

\bibitem[\protect\citeauthoryear{Cameir{\~a}o \bgroup \em et al.\egroup }{2009}]{gaming}
M{\'o}nica~S Cameir{\~a}o, Sergi Berm{\'u}dez~i Badia, Esther Duarte~Oller, and Paul~FMJ Verschure.
\newblock The rehabilitation gaming system: a review.
\newblock {\em Advanced Technologies in Rehabilitation}, pages 65--83, 2009.

\bibitem[\protect\citeauthoryear{Fisher \bgroup \em et al.\egroup }{2021}]{colmap}
Alex Fisher, Ricardo Cannizzaro, Madeleine Cochrane, Chatura Nagahawatte, and Jennifer~L Palmer.
\newblock Colmap: A memory-efficient occupancy grid mapping framework.
\newblock {\em Robotics and Autonomous Systems}, 142:103755, 2021.

\bibitem[\protect\citeauthoryear{Fridovich-Keil \bgroup \em et al.\egroup }{2023a}]{kplane}
Sara Fridovich-Keil, Giacomo Meanti, Frederik~Rahb{\ae}k Warburg, Benjamin Recht, and Angjoo Kanazawa.
\newblock K-planes: Explicit radiance fields in space, time, and appearance.
\newblock In {\em Proceedings of the IEEE/CVF Conference on Computer Vision and Pattern Recognition}, pages 12479--12488, 2023.

\bibitem[\protect\citeauthoryear{Fridovich-Keil \bgroup \em et al.\egroup }{2023b}]{kplanes}
Sara Fridovich-Keil, Giacomo Meanti, Frederik~Rahb{\ae}k Warburg, Benjamin Recht, and Angjoo Kanazawa.
\newblock K-planes: Explicit radiance fields in space, time, and appearance.
\newblock In {\em Proceedings of the IEEE/CVF Conference on Computer Vision and Pattern Recognition}, pages 12479--12488, 2023.

\bibitem[\protect\citeauthoryear{Kerbl \bgroup \em et al.\egroup }{2023}]{3DGS}
Bernhard Kerbl, Georgios Kopanas, Thomas Leimk{\"u}hler, and George Drettakis.
\newblock 3d gaussian splatting for real-time radiance field rendering.
\newblock {\em ACM Trans. Graph.}, 42(4):139--1, 2023.

\bibitem[\protect\citeauthoryear{Kingma and Ba}{2014}]{adam}
Diederik~P Kingma and Jimmy Ba.
\newblock Adam: A method for stochastic optimization.
\newblock {\em arXiv preprint arXiv:1412.6980}, 2014.

\bibitem[\protect\citeauthoryear{Liu \bgroup \em et al.\egroup }{2024}]{genn2n}
Xiangyue Liu, Han Xue, Kunming Luo, Ping Tan, and Li~Yi.
\newblock Genn2n: Generative nerf2nerf translation.
\newblock In {\em Proceedings of the IEEE/CVF Conference on Computer Vision and Pattern Recognition}, pages 5105--5114, 2024.

\bibitem[\protect\citeauthoryear{Luiten \bgroup \em et al.\egroup }{2024}]{dy3dgs}
Jonathon Luiten, Georgios Kopanas, Bastian Leibe, and Deva Ramanan.
\newblock Dynamic 3d gaussians: Tracking by persistent dynamic view synthesis.
\newblock In {\em 2024 International Conference on 3D Vision (3DV)}, pages 800--809. IEEE, 2024.

\bibitem[\protect\citeauthoryear{Machinery}{1950}]{embodiedAI}
Computing Machinery.
\newblock Computing machinery and intelligence-am turing.
\newblock {\em Mind}, 59(236):433, 1950.

\bibitem[\protect\citeauthoryear{Mildenhall \bgroup \em et al.\egroup }{2021}]{nerf}
Ben Mildenhall, Pratul~P Srinivasan, Matthew Tancik, Jonathan~T Barron, Ravi Ramamoorthi, and Ren Ng.
\newblock Nerf: Representing scenes as neural radiance fields for view synthesis.
\newblock {\em Communications of the ACM}, 65(1):99--106, 2021.

\bibitem[\protect\citeauthoryear{M{\"u}ller \bgroup \em et al.\egroup }{2022}]{ngp}
Thomas M{\"u}ller, Alex Evans, Christoph Schied, and Alexander Keller.
\newblock Instant neural graphics primitives with a multiresolution hash encoding.
\newblock {\em ACM transactions on graphics (TOG)}, 41(4):1--15, 2022.

\bibitem[\protect\citeauthoryear{Poole \bgroup \em et al.\egroup }{2022}]{dreamfusion}
Ben Poole, Ajay Jain, Jonathan~T Barron, and Ben Mildenhall.
\newblock Dreamfusion: Text-to-3d using 2d diffusion.
\newblock {\em arXiv preprint arXiv:2209.14988}, 2022.

\bibitem[\protect\citeauthoryear{Pumarola \bgroup \em et al.\egroup }{2021}]{dnerf}
Albert Pumarola, Enric Corona, Gerard Pons-Moll, and Francesc Moreno-Noguer.
\newblock D-nerf: Neural radiance fields for dynamic scenes.
\newblock In {\em Proceedings of the IEEE/CVF Conference on Computer Vision and Pattern Recognition}, pages 10318--10327, 2021.

\bibitem[\protect\citeauthoryear{Ravi \bgroup \em et al.\egroup }{2024a}]{sam2}
Nikhila Ravi, Valentin Gabeur, Yuan-Ting Hu, Ronghang Hu, Chaitanya Ryali, Tengyu Ma, Haitham Khedr, Roman R{\"a}dle, Chloe Rolland, Laura Gustafson, et~al.
\newblock Sam 2: Segment anything in images and videos.
\newblock {\em arXiv preprint arXiv:2408.00714}, 2024.

\bibitem[\protect\citeauthoryear{Ravi \bgroup \em et al.\egroup }{2024b}]{ravi2024sam2}
Nikhila Ravi, Valentin Gabeur, Yuan-Ting Hu, Ronghang Hu, Chaitanya Ryali, Tengyu Ma, Haitham Khedr, Roman R{\"a}dle, Chloe Rolland, Laura Gustafson, Eric Mintun, Junting Pan, Kalyan~Vasudev Alwala, Nicolas Carion, Chao-Yuan Wu, Ross Girshick, Piotr Doll{\'a}r, and Christoph Feichtenhofer.
\newblock Sam 2: Segment anything in images and videos.
\newblock {\em arXiv preprint arXiv:2408.00714}, 2024.

\bibitem[\protect\citeauthoryear{Wohlgenannt \bgroup \em et al.\egroup }{2020}]{vr}
Isabell Wohlgenannt, Alexander Simons, and Stefan Stieglitz.
\newblock Virtual reality.
\newblock {\em Business \& Information Systems Engineering}, 62:455--461, 2020.

\bibitem[\protect\citeauthoryear{Wu \bgroup \em et al.\egroup }{2024}]{4dgaussian}
Guanjun Wu, Taoran Yi, Jiemin Fang, Lingxi Xie, Xiaopeng Zhang, Wei Wei, Wenyu Liu, Qi~Tian, and Xinggang Wang.
\newblock 4d gaussian splatting for real-time dynamic scene rendering.
\newblock In {\em Proceedings of the IEEE/CVF Conference on Computer Vision and Pattern Recognition}, pages 20310--20320, 2024.

\end{thebibliography}

\end{document}